\documentclass[usegraphicx,useAMS,usenatbib]{mn2e}
\usepackage{times}
\usepackage{amssymb}
\usepackage{amsmath}
\usepackage{graphicx}
\usepackage{euscript}
\usepackage{rotating}
\usepackage{epsfig}
\usepackage{mathptmx}
\usepackage{amsbsy}

\def\approxlt{\mathrel{\spose{\lower 3pt\hbox{$\sim$}}
        \raise 2.0pt\hbox{$<$}}}
\def\approxgt{\mathrel{\spose{\lower 3pt\hbox{$\sim$}}
        \raise 2.0pt\hbox{$>$}}}

\def\arc{{\rm\thinspace arcsec}}
\def\cm{{\rm\thinspace cm}}

\def\erg{{\rm\thinspace erg}}

\def\ha{H$\alpha$}

\def\keV{{\rm\thinspace keV}}
\def\m{{\rm\thinspace m}}

\def\km{{\rm\thinspace km}}
\def\kpc{{\rm\thinspace kpc}}

\def\Mpc{{\rm\thinspace Mpc}}

\def\s{{\rm\thinspace s}}
\def\yr{{\rm\thinspace yr}}

\def\ergpcmsqpspsqarcsec{\hbox{$\erg\cm^{-2}\s^{-1}\arc^{-2}\,$}}

\def\ergps{\hbox{$\erg\s^{-1}\,$}}

\def\kmps{\hbox{$\km\s^{-1}\,$}}

\def\pcmsq{\hbox{$\cm^{-2}\,$}}

\def\psqarc{\hbox{$\arc^{-2}\,$}}

\def\sqarc{\hbox{$\arc^{2}\,$}}
\def\kmpspMpc{\hbox{$\kmps\Mpc^{-1}$}}

\DeclareMathAlphabet{\vib}{OML}{cmm}{m}{it} 
\begin{document}
\include{defn} 
\title{The extended H$\alpha$ emitting filaments surrounding 
NGC\,4696, the central galaxy of the Centaurus cluster}
\author[C. S. Crawford et al ] 
{C. S. Crawford, N. A. Hatch, A. C. Fabian and J. S. Sanders\\ 
Institute of Astronomy, Madingley Road,
Cambridge, CB3 0HA} 

\date{Accepted for publication in MNRAS 27 July 2005}

\maketitle

\begin{abstract} We present images of NGC\,4696, the central galaxy in
the Centaurus cluster, showing the large extent of cool filaments
which are bright in \ha\ line emission. These filaments share the
detailed structure of both the central dust lane and the inner regions
of the arc-like plumes seen in soft X-ray emission. The X-ray gas is
at its coolest, and most absorbed in this same region. The smoothness
of the features implies that the local environment is not strongly
turbulent.  We suggest that these filaments are either shaped by
confinement due to a strong magnetic field, or by bulk flows within
the intracluster medium.  We propose that like similar filamentary
systems in the core of clusters, these cooler components have been
drawn out of the central galaxy behind buoyant gas bubbles from
previous episodes of radio activity. We find a spur of low-frequency
radio emission leading to a region of low X-ray pressure within the
intracluster medium supporting this interpretation. 
\end{abstract}


\section{Introduction}

NGC\,4696 is the central dominant galaxy of the rich, nearby Centaurus
cluster of galaxies (A3526), lying at a redshift of $z=0.0104$.
Line-emitting filaments were first discovered to extend out to the
south-west of the galaxy over twenty years ago by Fabian et al (1982).
NGC\,4696 has also been long-known to harbour a distinctive dust lane
sweeping around the core of the galaxy to the south and west
(Shobbrook 1966; Jorgensen et al 1983; Sparks, Macchetto \& Golombek 1989; Laine et al
2003).  There is a close spatial correspondence between the structure
in the extended line emission and the dust lane (Sparks et al 1989).
Neutral gas has been detected as sodium D in absorption against the
galaxy stellar continuum, and is assumed to be associated with the
emission line system and dust lane (Sparks, Carollo, Macchetto 1997).
High-resolution optical (HST) images show that the galaxy has a double
morphology within the central arcsecond, presumably due to the active
nucleus (Laine et al 2003). 



NGC\,4696 hosts the small steep-spectrum FR~I-type radio source
PKS~1246--410, which has a morphology distorted by confinement by the
surrounding intracluster medium.  The two radio lobes to the east and
west of the core are compact, and sweep down to the south; they have
excavated small cavities in the X-ray gas (Taylor, Fabian \& Allen
2002; Taylor et al 2005).  

The X-ray emission from the surrounding cluster is consistent with 
cooling flow of around ten solar masses a year (Sanders \& Fabian
2002; Allen et al 2001; Ikebe et al 1999).  Within an arcminute of the
central cluster galaxy, the X-ray emission is disturbed, showing not
only the radio cavities but also excess soft emission in the form of
plume-like features that sweep from the south of the central cluster
galaxy to trail out and off to the north-east (Sanders \& Fabian 2002,
Fabian et al 2005). The origin of these features is not known; the gas
within the plumes is the same metallicity as, but cooler than the
surrounding intra-cluster medium.

In this paper we present ground-based data and archival HST data with
the aim of examining in detail the relationship between the dust lane,
the optically-emitting filaments, the soft X-ray plumes, X-ray
absorbing gas and central radio source. Throughout the paper we assume
the cosmology of $H_0=71$\kmpspMpc, $\Omega_m=0.7$ and
$\Omega_\Lambda=0.3$, for which 1 arcsec corresponds to 0.21\kpc\ at
the distance of 44.3~Mpc for NGC\,4696. 

\section{X-ray data}
X-ray analysis was undertaken using the 199~ksec {\sl Chandra}
Centaurus dataset of Fabian et al (2005).  The complicated structures
in the soft X-ray emission (the radio cavities and the plumes), and the
X-ray temperature map from these data are shown for reference
in the first panels of Figs~\ref{fig:panels} and \ref{fig:extin}. 

Regions containing $\sim 400$~counts between 0.5 and 7 keV were
selected, using the bin accretion tessellation algorithm of Cappellari
\& Copin (2003). Spectra were extracted from each of the regions from
each of the datasets, and were added together to form a total
spectrum.  
A blank sky spectrum was obtained from a blank sky observation,
tailored to the Centaurus observations, and extracted from a similar
region of CCD to that of the regions examined. 
Responses and ancillary responses were generated for each of the
regions and datasets using the standard CIAO \textsc{mfwarf} and
\textsc{mkacisrmf} tools. The responses and ancillary responses for
each of the datasets for a particular region were then averaged
together, weighted according to the fraction of counts in a particular
dataset for that region. The spectra were fit between 0.5 and 7~keV
using a \textsc{mekal} emission model (Mewe, Gronenschild \& van den
Oord 1985; Liedahl, Osterheld \& Goldstein 1995) absorbed with a
\textsc{phabs} model (Balucinska-Church \& McCammon 1992), minimising
the C statistic (Cash 1979). 

Values of the column density (N$_{\rm H}$) inferred vary from the
Galactic column (around $8\times10^{20}$ \pcmsq) up to
5$\times10^{21}$ \pcmsq; the resulting column density distribution in
the X-ray gas around NGC\,4696 is shown in Fig~\ref{fig:extin}. The
statistical errors are around 5-10 per cent in regions where the
N$_{\rm H}$ is high, although there may be further errors from the
approximation of fitting single temperature models in a region where
there is evidence for multi-temperature gas. 
Although we applied a correction to the ancillary responses for the
build up of contaminant at low energies on the ACIS-S3 detector, there
may still be residual systematic uncertainties in the calibration here.
We expect the relative trends in $N_\mathrm{H}$ to
be correct, even if the absolute normalisation is not. In addition we
are concentrating on data taken from the centre of the CCD, well away
from the edges where the contaminant is more significant. 

An analogue to the X-ray thermal gas pressure in the intracluster
medium was created by multiplying the X-ray temperature as a function
of position by the square root of the emission measure per unit area.
These were measured by fitting a \textsc{mekal} spectral model to
X-ray spectra extracted from spatial regions. The spatial regions were
chosen to contain approximately $10^4$ counts in total using a contour
binning algorithm (Sanders in preparation). The method takes a
smoothed X-ray image and chooses bins which group together pixels of
similar smoothed flux. Here the raw X-ray image was smoothed using a
top-hat circular kernel with a size which varied to ensure there was a
signal to noise of 30 within the kernel, or over 900 counts. Starting
from the brightest smoothed pixel, neighbouring pixels closest in
smoothed flux to the original pixel are added to a bin until the
signal to noise calculated from the unsmoothed data exceeds a
threshold value (here 100). When this threshold is reached a new bin
is created. An additional geometric criterion is used to ensure the
bins are not too elongated. A pixel is not added if it lies at a
distance of greater than 1.9 times the radius of a circle with the
same area as the bin from the bin centroid. In order to look at
deviations from symmetry within the X-ray gas surrounding NGC\,4696,
we calculated the mean value of the pressure analogue at each radius,
and subtracted it from the pixels at that radius. The resulting map of
relative thermal pressure are discussed later in context in section 6. 

\section{Ground-based Optical Observations}

Optical observations of NGC\,4696 were taken with the ESO Multi-Mode
Instrument (EMMI) on the 3.58\m\ New Technology Telescope at La Silla,
Chile on the nights of 2004 April 11 and 2004 May 20.  Images were
taken through the broadband $I$ (ESO \#610) and $B$ (ESO \#603)
filters, and a narrowband H$\alpha$ filter centred on
$\lambda=6631\mathring{\rm{A}}$ with a full-width at half-maximum of
66$\mathring{\rm{A}}$ (ESO \#597).  The narrow-band filter covers the
wavelength of the H$\alpha$ emission line at the redshift of
NGC\,4696, but only allows in around half of the redshifted [N{\sc
ii}]$\lambda6584$ line emission (although all of the weaker [N{\sc
ii}]$\lambda6548$ line). Assuming a ratio of 
[NII]$\lambda\lambda6548,6584$/\ha\ of 3.2 (consistent with Johnstone,
Fabian \& Nulsen 1987 and Lewis, Eracleous \& Sambruna
2003), 
we obtain \ha-only fluxes assuming by dividing the observed
 (continuum-subtracted) line fluxes by three. 
The seeing conditions were constant at
around 0.9~arcsec for both the $I$- and \ha-band observations, and at around
1.2~arcsec for the $B$-band observation. 

The $I$-band (taken as 5$\times$330\,s exposures) and H$\alpha$ images
(7$\times$700\,s+1$\times$680\,s) were observed using the Red Imaging
and Low Dispersion Spectroscopy (RILD) mode of EMMI, which has a field
of view of $9.1\times9.9$ square arcminutes.  
The EMMI Red 2K$\times$4K CCD detector has a pixel size of 
$0.333\times0.333$\sqarc. The $B$-band images (taken as
2$\times$1100\,s+1$\times$1050\,s exposures) were observed using the
Blue Imaging Mode (BIMG), which has a $6.2\times6.2$ square arcminute
field of view, and a 1K$\times$1K Blue CCD detector with a pixel
size of $0.37\times0.37$\psqarc. 
Multiple bias frames and dome flat-fields 
were taken on each of the observing nights,
as well as the Landolt standard star field PG1525 in all three
bands, and the spectroscopic standards LTT\,7379 and LTT\,6248, in the
$I$-band and narrow-band H$\alpha$ only. A summary of the observations is
given in Table~\ref{tab:log}.

\begin{table}
\begin{tabular}{|lllrl|}
Target&Filter &Seeing & Total Integration & pixel size\\
&(ESO filter number)&arcsec&time (seconds)&(arcsec)\\ 
NGC\,4696&I (ESO \#610)&0.9&1650&0.333\\
NGC\,4696&H$\alpha$ (ESO \#597)&0.9&5580&0.333\\
NGC\,4696&B (ESO \#603)&1.15&3250&0.37\\
\end{tabular}
\caption{Summary of ESO optical observations}
\label{tab:log}
\end{table}

\subsection{Data Reduction }

The observations were bias-subtracted, flat-fielded, and normalised to
a 1 second exposure time. The sky background was then removed from all
NGC\,4696 images by subtracting the mode of the background pixels
furthest from the galaxy.

In order to flux-calibrate a galaxy-subtracted image of the H$\alpha$
filaments, both the H$\alpha$ and $I$-band images were flux-calibrated
from observations of the spectroscopic standards LTT\,7379 and
LTT\,6248; the reference spectra of these standards are provided by
Hamuy et al (1992).  The observed spectra of the flux standard stars
were convolved with the filter transmission functions to obtain an
average calibrated flux of 1.25$\times10^{-16}$ erg s$^{-1}$ cm$^{-2}$
(counts s$^{-1}$)$^{-1}$ for the H$\alpha$ image and
1.09$\times10^{-16}$erg s$^{-1}$ cm$^{-2}$ (counts s$^{-1}$)$^{-1}$ for the $I$-band image.
The $I$-band image was divided by 35.9 to account for the difference
between the $I$-band and H$\alpha$ filter transmission functions. The
continuum was removed from the H$\alpha$ image by subtracting the
normalised $I$-band image. Finally, the image was corrected for
Galactic extinction assuming a Galactic E(B-V)=0.115 (A$_{B}$=0.46)
(Burstein \& Heiles 1984) which increased the observed H$\alpha$ flux
by 33 percent. 

\begin{figure*}
\centering
\includegraphics[width=2\columnwidth]{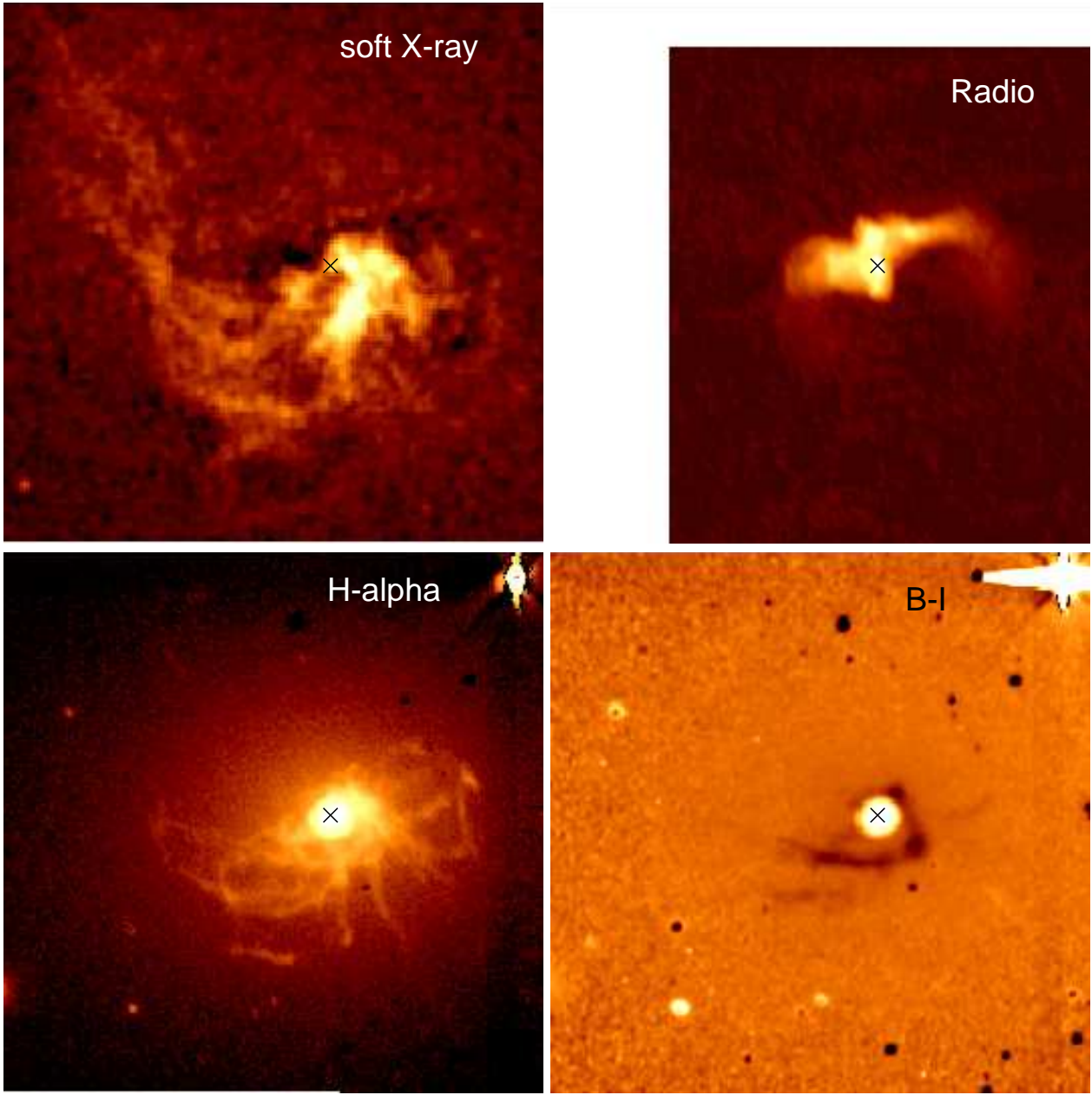}
\caption{ 
Clockwise from top left: 
the low-energy (0.3-1\keV), unsharp-masked  {\sl Chandra} X-ray image; 
the 5-GHz radio intensity image (the data are from Taylor et al 2002); 
the Gemini $B-I$ image (where regions 
of high extinction are coloured dark, and 
the central region appears white
because the core is saturated in the $I$-band image); 
and the continuum-subtracted \ha\ image
of NGC\,4696. All images are aligned and matched in scale. 
North is to the top and East to the left, and individual panels are 
each 83 arcsec on a side. 
The black cross in all the images marks the location of the radio source
core for reference. }
\label{fig:panels}
\end{figure*}

A $B-I$ colour image was also created to map out the distribution of
the absorbing dust.  The difference in pixel sizes required the
$B$-band image to be rebinned using the IRAF \lq drizzle' program in
order to make it compatible with the $I$-band image.  The $I$-band
image was divided by 2.4 to account for the difference between the
$I$- and $B$-band filter transmission functions. To calibrate this
image, both the $B$- and $I$-band filter images were flux-calibrated
using the Landolt Standard field PG1525.  Inspection of the $B-I$
colour image suggests that the Northeast quadrant of the galaxy is the
least affected by dust absorption (although not entirely unaffected).
We thus assume the intrinsic colour of the underlying galaxy can be
best estimated as the average colour of the galaxy in this region, and
subtract this from the individual $B$- and $I$-band images. 
The final continuum-subtracted \ha\ image is shown along with the 
$B-I$ colour map in Fig~\ref{fig:panels}. 

\subsection{Dust absorption analysis } 

\begin{figure*}
 \centering
\hbox{
\includegraphics[width=0.67\columnwidth]{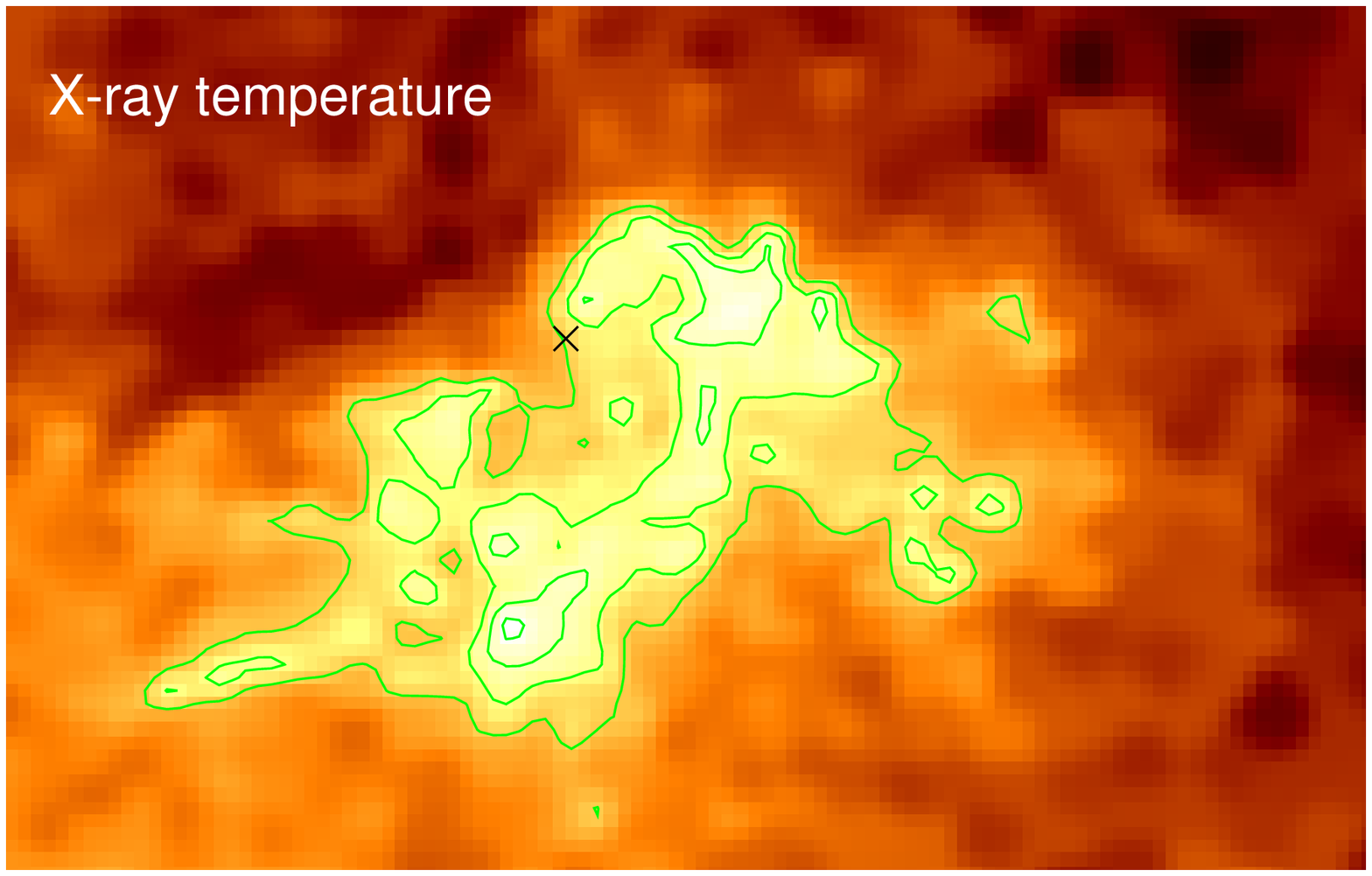}
\includegraphics[width=0.67\columnwidth]{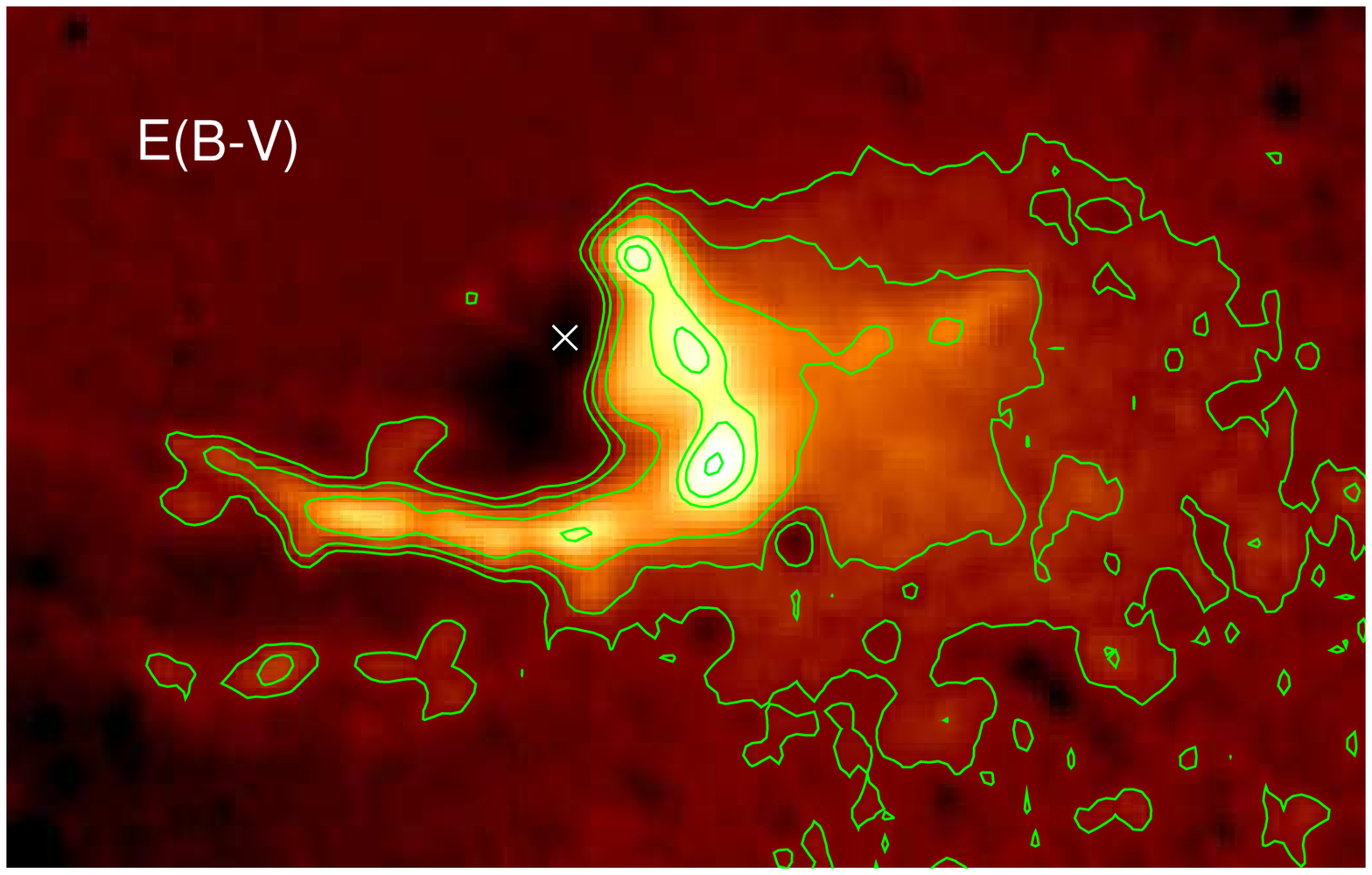}
\includegraphics[width=0.67\columnwidth]{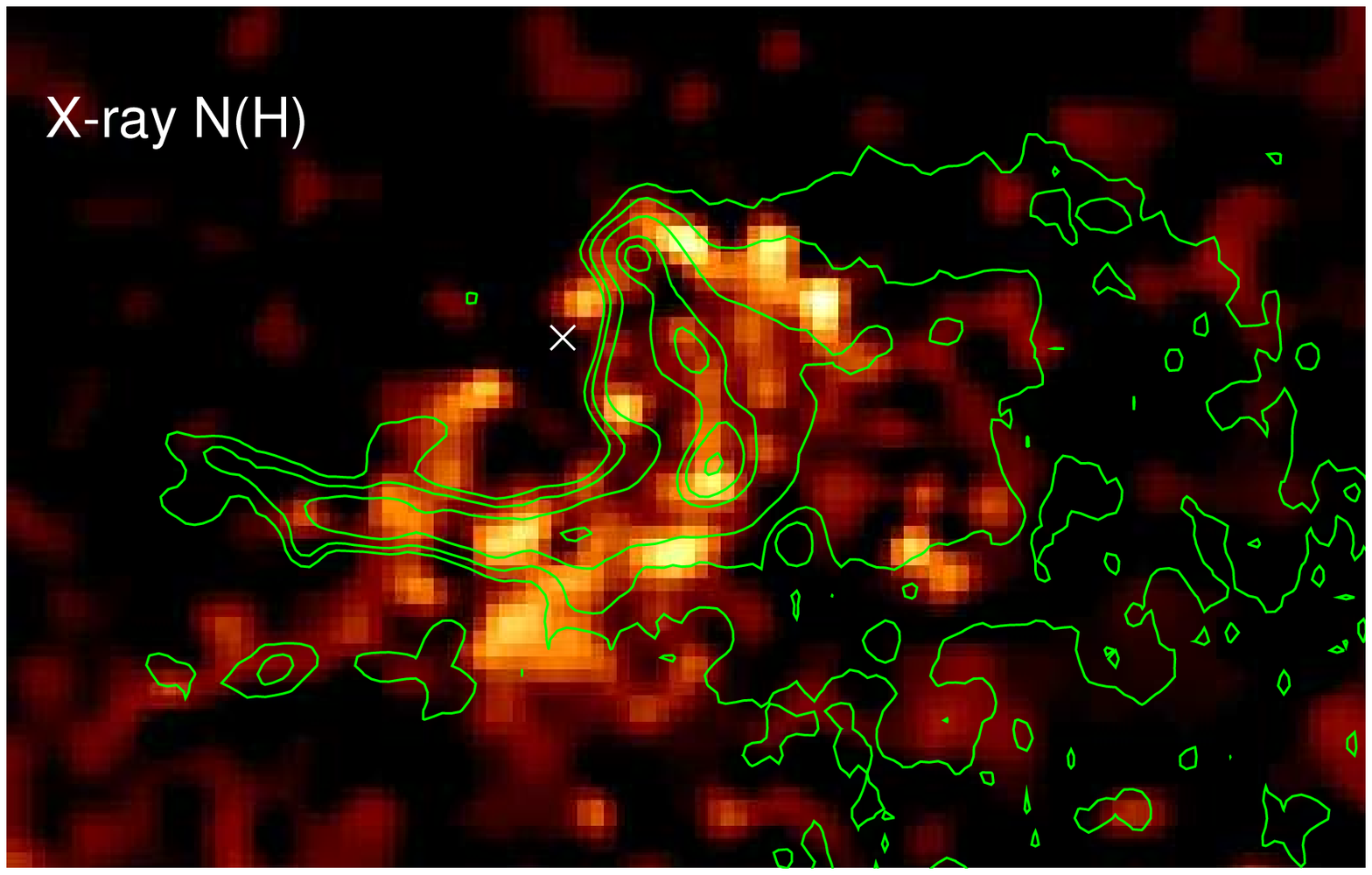}
}
\caption{ 
The X-ray temperature (left: with contour levels at 0.56, 0.58, 0.65,
0.75 and 0.9 in keV -- but note that the innermost contours are the
lowest temperatures);  $B$-band extinction map (middle; contours at 0.01, 0.02,
0.04, 0.08, 0.1, 0.14 in E(B-V)); 
X-ray column density (right: with the contours from the $B$ band
extinction map superposed). The lighter regions in the 
X-ray N$_{\rm H}$ column density map represent absorption columns of
$2-5\times10^{21}$\pcmsq. 
The images are scaled to match each other, with each image showing 
$\sim50\times$30 arcsec. The cross marks the position of the radio
core in all panels. }
\label{fig:extin}
\end{figure*}

In order to determine the absolute levels of extinction we create an
intrinsic galaxy model, assuming it follows the galaxy profile in the
north-east quadrant and extending this into a azimuthally symmetric
form for each of the $I$ and $B$-band images. The model galaxy was
Gaussian smoothed by two pixels and then subtracted from the $I$- and
$B$-band images to create $A_{B}$ and $A_{I}$ extinction maps.  These
were transformed to maps of E(B-V) (eg Fig~\ref{fig:extin})
assuming $A_{B}=4.354E(B-V)$, $A_{I}=2.0E(B-V)$ (Zombeck 1992) 
 
Due to the higher sensitivity of the $B$-band to extinction, and the 
fact that the $I$-band image is affected by saturation of the bright
nuclear regions, we take the $B$ image as the more accurate
representation of the extinction in the galaxy. Table \ref{tab:ex}
gives a quantitative description of the information from the $B$--band
image in Fig~\ref{fig:extin}.  The estimated
column density, $N$(HI+H$_2$) assumes the same relation between column density
and reddening as for our own Galaxy, as given by Bohlin, Savage, and
Drake (1978), 
as $ N$(HI+H$_2$)$=5.8 \times 10^{21} E(B-V)$\pcmsq. 

\begin{table}
\begin{tabular}{|crr|}
Extinction&Area &Estimate Hydrogen \\
&(arcsec$^{2}$) & Column Density (\pcmsq)\\ 
0.01$<E(B-V)<$0.02& 310&8.5$\times10^{19}$\\
0.02$<E(B-V)<$0.04& 135&1.7$\times10^{20}$\\
0.04$<E(B-V)<$0.08& 65&3.5$\times10^{20}$\\
0.08$<E(B-V)<$0.1& 14&5.2$\times10^{20}$\\
0.1$<E(B-V)<$0.14& 7&6.9$\times10^{20}$\\
\end{tabular}
\caption{Quantitative values for the dust extinction within each of the
contour levels of Fig~\ref{fig:extin}, as taken from the $B$-band
image only.  }
\label{tab:ex}
\end{table}

\section{Archival HST Observations}

Three datasets of NGC\,4696 were extracted from the HST archive. These
are taken through the F555W (WFPC2 $V$-band), F702W (WFPC $R$-band)
and F814W (WFPC $I$-band) filters; the observations used are shown in
Table~\ref{tab:loghst}. The dust lane is visible in all the data, and
has been briefly remarked upon previously by Laine et al (2003). We
attempt here to obtain a combined image that shows the dust lane
structure in much more detail.  The two raw frames in each band were
co-added to remove cosmic rays, and the three resulting images summed
to form an all-band image (after appropriate shifting of the F814W
data to match the other datasets). 

\begin{figure*}
\centering
\hbox{
\includegraphics[width=0.67\columnwidth]{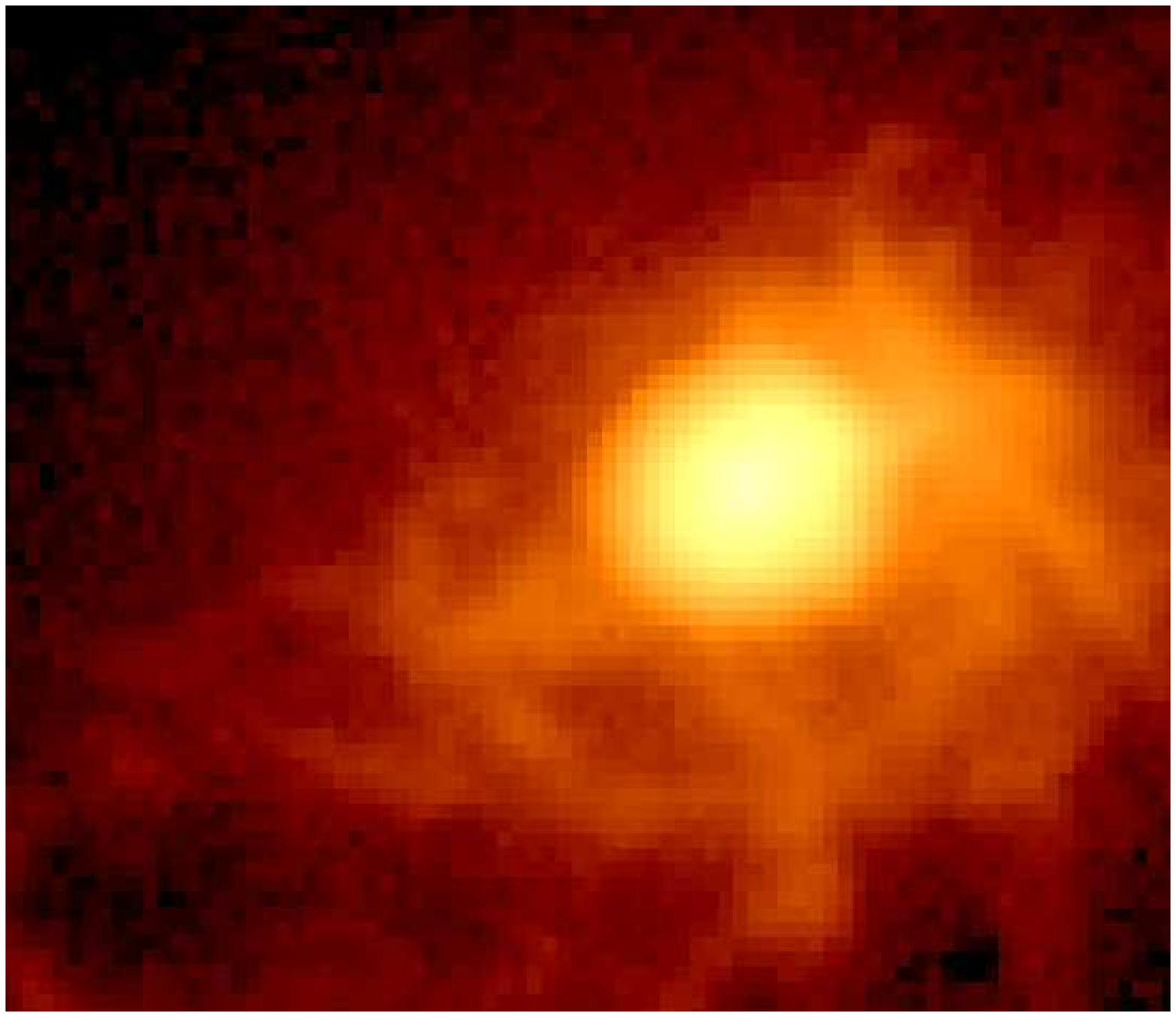}
\includegraphics[width=0.67\columnwidth]{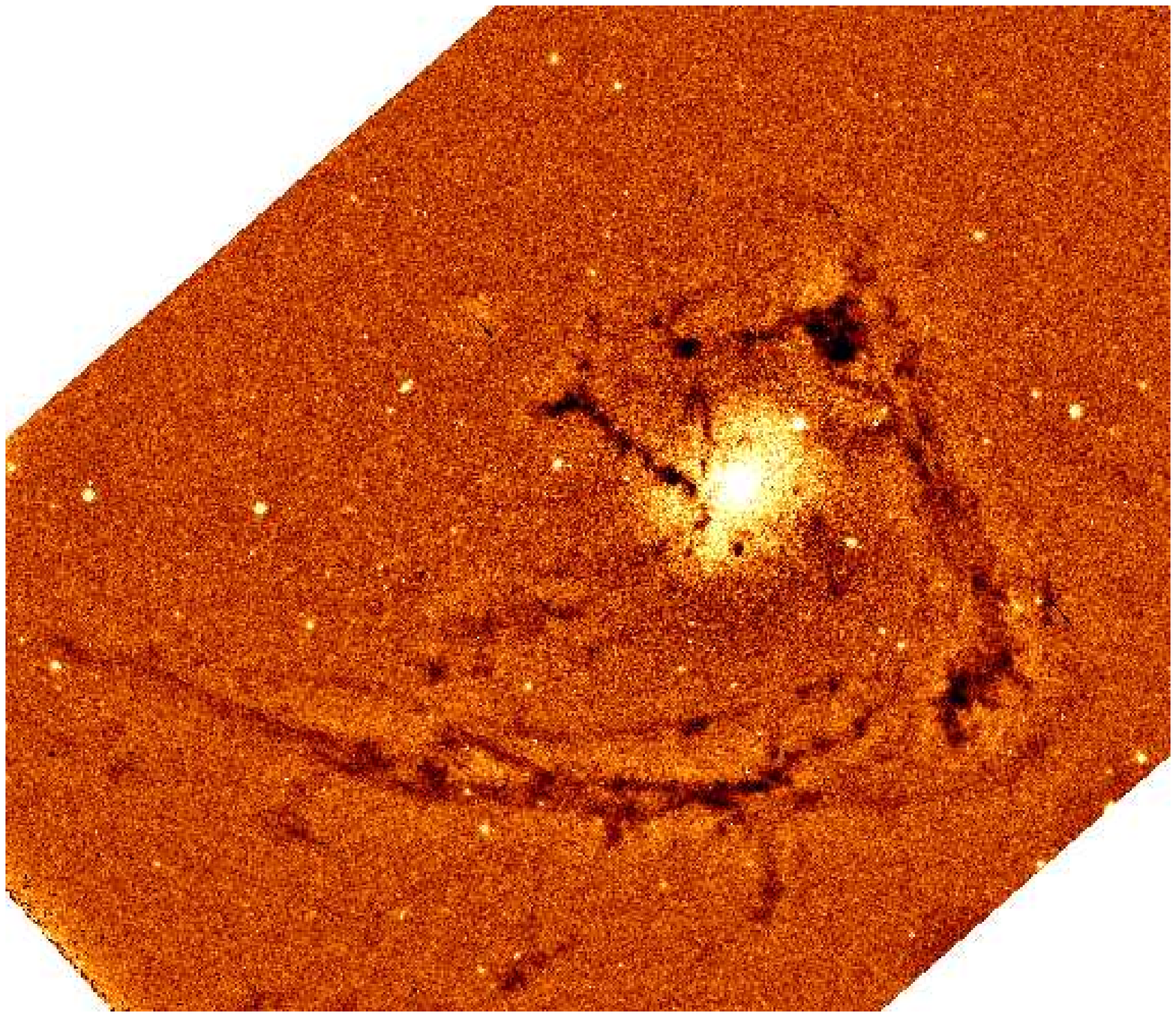}
\includegraphics[width=0.67\columnwidth]{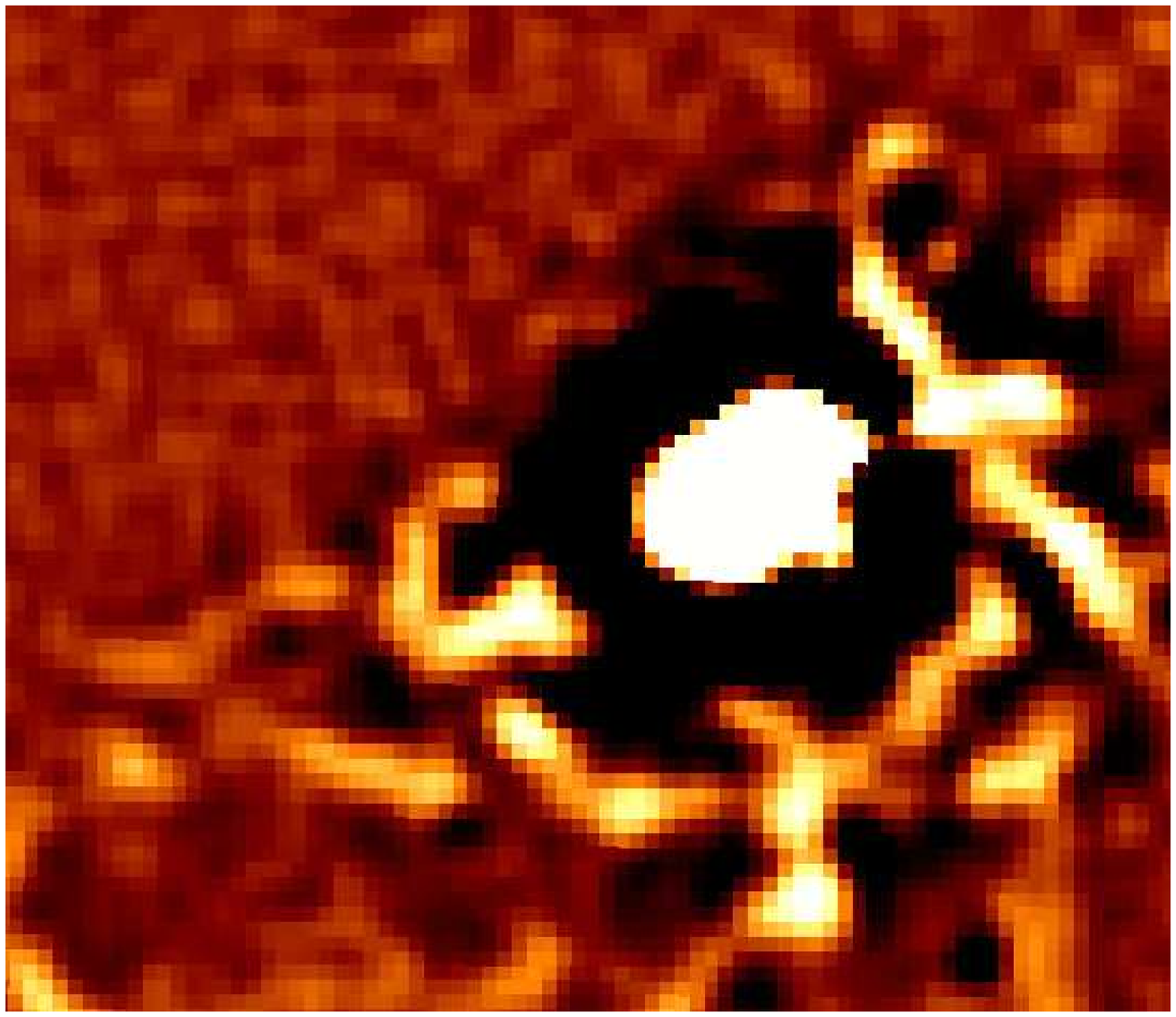}
}
\caption{ 
The (continuum-subtracted) \ha\ emission from the ground-based
Gemini observations (left); the
unsharp-masked image of the dust lanes from the HST data (middle) and
an unsharp-masked version of the \ha\ image (right). The images are
aligned on 
matched scales, and each image is $\sim26$ arcsec $\times23$ arcsec.} 
\label{fig:haandhstdust}
\end{figure*}

The summed image was unsharp-masked in order to reveal the dust lane,
by subtracting from this image a version of itself Gaussian-smoothed
by 13 pixels (corresponding to $\sim$0.6~arcsec).  The unsharp-masked
image shows considerable further structure in the dust absorption
features at high resolution and is shown in
Figure~\ref{fig:haandhstdust}. 

Our combined HST image confirms the double nucleus in the core of the
galaxy (Laine et al 2003), with two components separated by 0.3
arcsecond along the north-east:south-west direction, and the brighter
to the north-east. 

\begin{table}
\begin{tabular}{crl}
Name & Exposure & Filter \\
 & (sec) & \\
U3560101B & 320 & F555W \\
U3560103B & 320 & F702W \\
U62G8401B & 1000 & F814W \\
\end{tabular}
\caption {Archival HST observations used.}
\label{tab:loghst}
\end{table}

\section{Results}

\subsection{Absorbing dust and gas}

Both the EMMI (Fig~\ref{fig:panels}) and HST
(Fig~\ref{fig:haandhstdust}) data show the primary dust lane structure
in the form of the \lq horseshoe' shape extending anti-clockwise from
the south-east round to the north-west, as described by Jorgensen et
al (1983). Both datasets also clearly show fainter, further structures
such as a secondary lane to the south-east, about twice as far from
the nucleus as the main dust lane and a radial lane to the west, again
beyond the main dust lane structure. Such features were hinted at in
the image of Sparks et al (1989). 

However, the more detailed structure in the HST image now shows the
horseshoe shape to appear more as a full spiral around the core,
continuing further round the north and into a radial feature extending
from the north-east into the centre of the galaxy at around 60 degrees
east of north -- the lane almost entirely encircles the nucleus
(Fig~\ref{fig:haandhstdust}).  There are further radial structures
near the core of the galaxy, at 20 and 120 degrees east of north (the
latter is also apparent in the EMMI image).  The HST data clearly resolve
the main dust lane into an elongated network of several thin (ie
spatially unresolved in the HST) filaments. 

Our highest values for the extinction lie in the three prominent knots
along the main dust lane, each with $E(B-V)>0.1$, in good agreement with 
estimates obtained by Jorgensen et al (1983), Sparks et al (1989) and
Wills et al (2004).
 
The maps of the X-ray absorption column density and temperature
(Fig~\ref{fig:extin}) reveal that N$_{\rm H}$ is highest, and the
X-ray gas is coolest in a region that is a similar in extent and shape
to that encompassed by the dust lane.  The positions of discrete blobs
of high extinction within the dust lane, however, do not seem to
correspond in detail with the highest regions of X-ray absorption
(right hand panel of Fig~\ref{fig:extin}). 
Typical X-ray temperatures are 0.6-0.7\keV\ and column densities in
this region vary between 2-5$\times10^{21}$\pcmsq.  Given the
foreground X-ray absorption due to our Galaxy in the direction of
Centaurus is $8\times10^{20}$\pcmsq, the values of excess $N$(H)
compare very well to those predicted from the optically-derived E(B-V) in
Table~\ref{tab:ex}. This suggests that the dust-to-gas ratio in this
region is not very different from that in our Galaxy. 

\subsection{The structure of the \ha\ filaments}

The narrow-band imaging resolves the emission-line nebula into a
complex network of filaments extending primarily from the south-east
through south to the west of the galaxy. At its widest, the nebula is
approximately 1 arcminute across (corresponding to just over 12\kpc).
The width of the \ha\ filaments is, at 0.9 arcsecond, spatially
unresolved ($<$0.2\kpc). 

Filaments within 8~arcsecond of the radio core have a typical surface
brightness of 3.5$\times 10^{-16}$\ergpcmsqpspsqarcsec, whereas the
outer filaments that stretch to over 27~arcsecond distant are fainter
at 1.4$\times 10^{-16}$\ergpcmsqpspsqarcsec.  We correct for the
contamination by the [NII] doublet in our filter, but an accurate
estimate of the total \ha\ luminosity of the entire nebula is complicated by
the saturation of the central regions in the $I$-band image; a lower 
limit is obtained by excluding the central saturated regions
to give a total \ha\ luminosity beyond a radius of 3.5 arcsec
(0.75\kpc) of 
L$_{\rm H\alpha}>$1.5$\times10^{40}$\ergps. The intrinsic luminosity 
(ie that not absorbed by the dust lanes) will be higher. 

\begin{figure*}
 \centering
\hbox{
  \includegraphics[width=1\columnwidth]{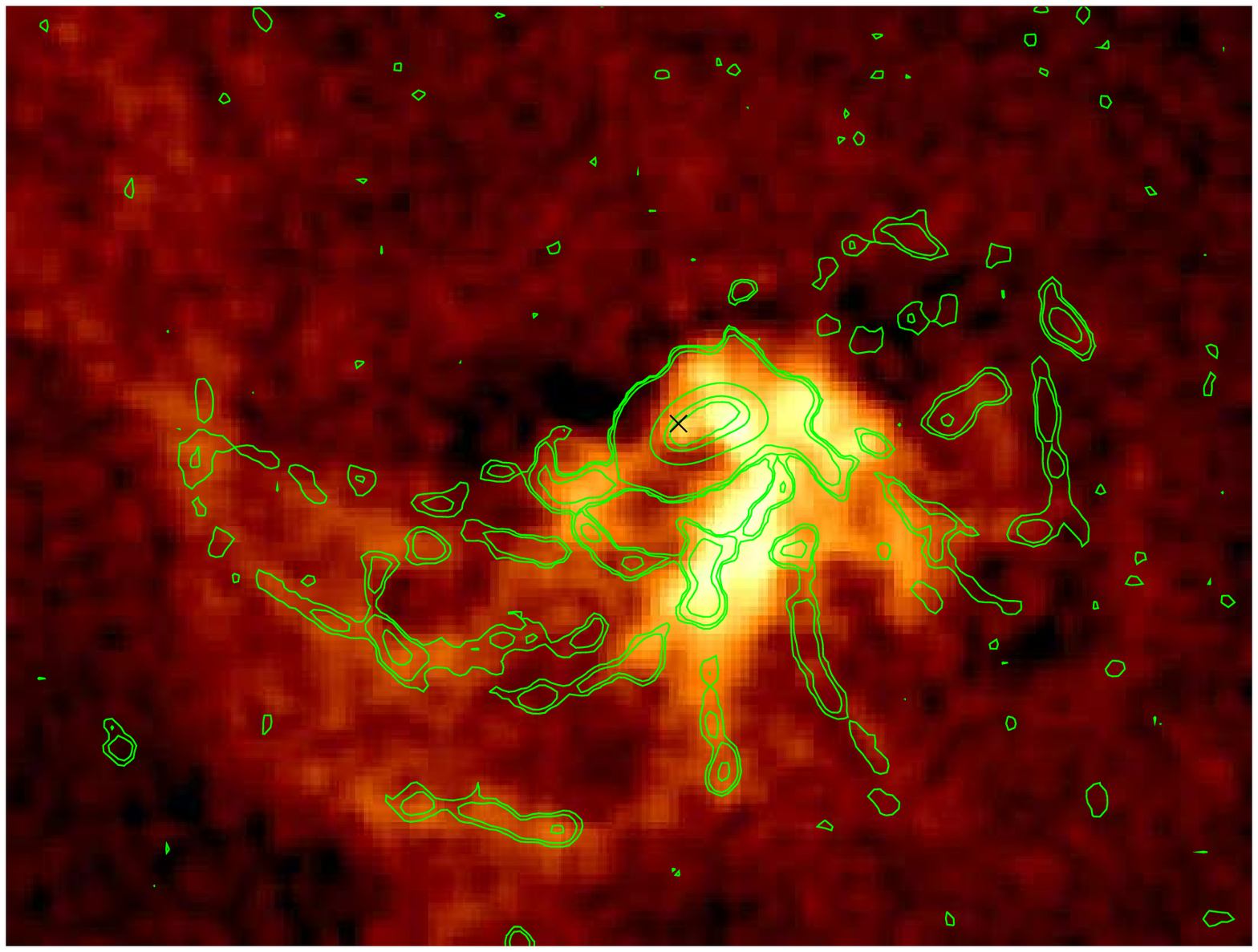}
  \includegraphics[width=1\columnwidth]{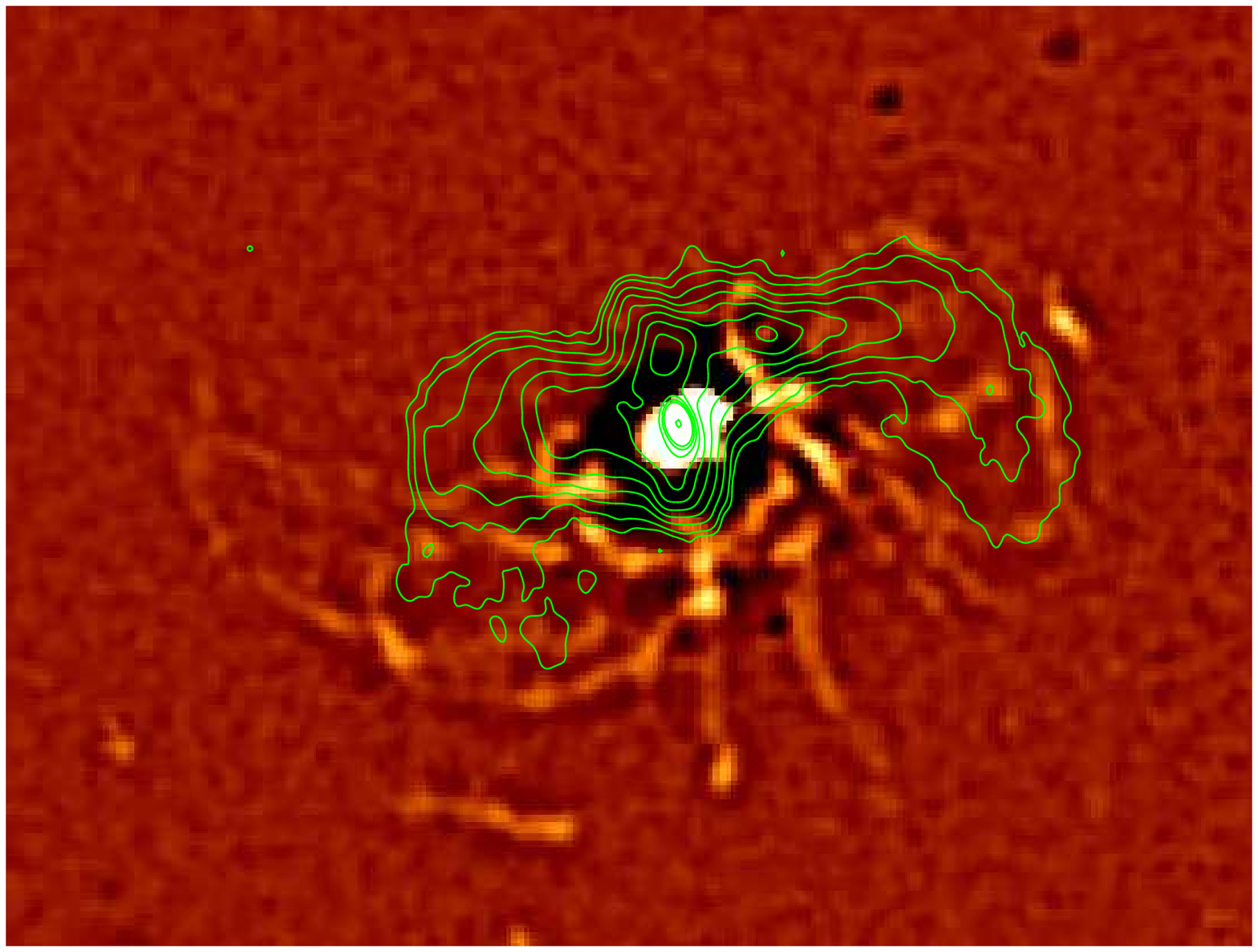}
  }
\caption{ 
Left:  soft X-ray image of NGC\,4696 with contours of the \ha\ emission
superposed (contour levels at 0.6, 1.5 and
4.5$\times10^{-17}$\ergpcmsqpspsqarcsec). \newline
Right: 
Contours of the 5~GHz radio emission 
superposed on the unsharp-masked \ha\ line emission image of NGC\,4696. 
The images are 68$\times$52 arcseconds
The black cross in the left hand panel marks the position of the 
radio source core. } 
\label{fig:contim}
\end{figure*}

There is a clear and close spatial correspondence between the
emission-line filaments and the inner regions of the long plumes of
soft X-ray emission (Figs~\ref{fig:panels} and \ref{fig:contim}). 
They directly overlap where the radial filaments stretch south of the
galaxy and where the three main filaments sweep to the south-east. The
\ha\ filaments also seem to encircle the western radio lobe, outlining
the X-ray cavity there (Fig~\ref{fig:contim}). Apart from this region,
most of the X-ray and \ha\ structures lie south of the diffuse radio
emission. 

The inner \ha\ filaments also directly trace the structures of the
dust lane as seen in the HST and ground-based images
(Figs~\ref{fig:panels} and \ref{fig:haandhstdust}), including such
details as the radial spur at 120 degrees east of north, the faint
western structure, and the fainter more southerly dust arm. The \ha\
also is encompassed in the coolest regions of the X-ray emission, and
highest regions of X-ray-inferred column density. 

\begin{figure}
\centering
\includegraphics[width=1\columnwidth]{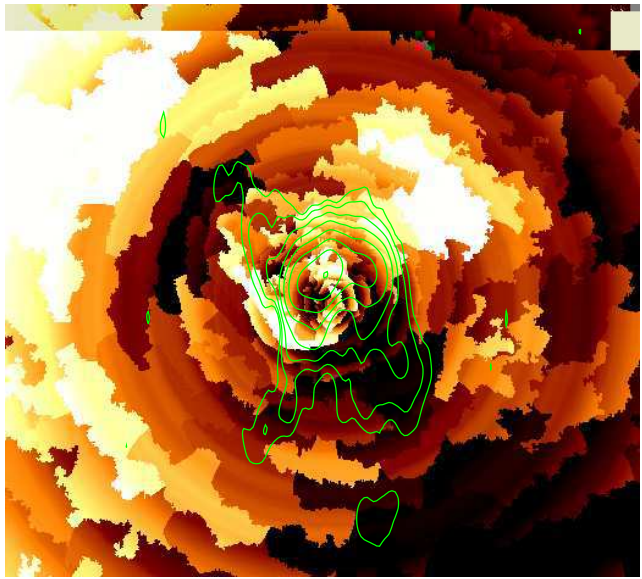}
\caption{ 
Map of the relative X-ray-inferred thermal pressure in the intracluster medium of the
Centaurus cluster, showing the deficit of pressure located at the end
of the soft X-ray plume to the north-east of NGC\,4696. 
The pressures are scaled relative to the mean pressure of the
intracluster medium at that radius, with deficits of --13 per cent
showing as black, and excesses above $+$14 per cent showing as white. 
The 330~MHz radio contours (from Fabian et al 2005) are superposed in
green, starting at 20~mJy/beam and increasing by factors of two. 
The image is $\sim3.9$ arcmin $\times3.6$ arcmin.} 
\label{fig:pmap}
\end{figure}

\section{Discussion} 

The arc structures seen in both X-ray and the optical line emission 
are smooth, and like those seen in Perseus and A\,1795 (Conselice et
al 2001; Fabian et al
2003; Fabian et al 2001; Crawford, Sanders \& Fabian 2005) show no
indication on small (kiloparsec and less) scales of surrounding
turbulence.  This smoothness, and the fact that the dust lanes,
emission-line and X-ray gas all share the same structure, suggests
that these features have been shaped either by a strong magnetic
field, or a bulk (and locally laminar) flow within the intracluster
medium.  Assuming that the filaments lie on the outside of the radio
cavities (Fig~\ref{fig:contim}), deeper and higher resolution radio
rotation measurements over the region where the radio source and the
filaments coincide spatially could quantify the strength of the
magnetic field within the filaments. 

If the filaments are all shaped by an external bulk flow, the way they
appear to spiral out from NGC\,4696 suggests that all these
components have been dragged out from the central galaxy; the spiral
appearance could be exaggerated if the plume is directed towards us. 
This is also supported by the fact that the densest, and coolest X-ray
gas is located in this region (Fig~\ref{fig:extin}). This extraction
could be in the wake of a buoyant bubble from a past episode of radio
activity in the galaxy. Examples of clusters where this phenomenon is
well documented include M87 in Virgo (B\"ohringer et al 1995; Churazov
et al 2001) and NGC\,1275 in the Perseus cluster (Fabian et al 2003).
The direction of the filaments would suggest that the past or \lq
ghost' bubble should lie to the north-east of the galaxy.  There is a
spur of low-frequency radio emission in this direction, ending at a
noticeable deficit (at around --6 to --13 per cent) 
in a map of the X-ray-inferred thermal pressure in the
intracluster medium (Fig~\ref{fig:pmap}). A very marginal deficit in
X-ray intensity is seen in the same location.  The pressure deficit is
centred approximately at RA 12:48:54.0 DEC --41:18:13 J2000), has the
thin linear morphology expected from a rising cap bubble, and is
tangential to the cluster core. Despite showing up as a sharp dip in
the X-ray pressure map it is not apparent in the X-ray temperature map,
suggesting the deficit in X-ray emission is due to a true drop in
density. Figure~\ref{fig:pmap} also shows that the low-frequency
radio emission to the south and south-west of the central cluster
galaxy also completely occupies a similar pressure deficit. 

The curve of the features (which may be partially due to projection
effects) suggests that the bubble has not risen radially. This may be
due to the central galaxy not being entirely in dynamical equilibrium
with the cluster potential, but still moving around the core (a
relative line-of-sight motion of 100-200\kmps\ is possible: Sanders \&
Fabian 2002).  The way the filaments appear to be pushed southward by
the current radio emission (Fig~\ref{fig:contim}) also points to their 
origin being within the central galaxy. 

Although on a much smaller spatial scale, and of considerably lower
luminosity, the \ha\ and X-ray filaments around NGC\,4696 are very
similar to those associated with the central galaxies in the Perseus
and A\,1795 clusters of galaxies. It is thus likely, by comparison to
the filaments around Perseus (Hatch et al 2004) that the filaments
around Centaurus also contain molecular hydrogen. The source of the
excitation and ionization of such filaments is a matter of
contraversy, with star formation being one possibility among many (see
discussion in Crawford et al 1999, for example). 

The X-ray filaments appear broader than the corresponding
line-emitting components. We have measured the flux of the bright
filament 22 arcsec to the SE of the nucleus. Its surface brightness is
about $7\times 10^{-15}$\ergpcmsqpspsqarcsec in X-rays (0.1--10\keV,
corrected for absorption) and $1.4\times 10^{-16}$\ergpcmsqpspsqarcsec
in H$\alpha$ (not corrected for intrinsic absorption).  The total
surface brightness implied by recombination, much of which emerges in
the UV, is about 20 times the \ha\ brightness, which is still only
about half the X-ray value. The filaments in the Centaurus cluster are
therefore X-ray bright. In contrast, those in the Perseus cluster
(Fabian et al 2003b) are UV bright (with a recombination flux brighter
than the X-ray flux by up to two orders of magnitude). This may in
part be due to the apparent high dust content of the Centaurus
filaments suppressing the observed \ha\ flux.  A simple equilibrium
conduction model applied to the X-ray data from the filaments in the
two different clusters implies that conductivity must be suppressed
well below the Spitzer rate in the Perseus cluster (Fabian et al
2003b) but would need to be enhanced above the Spitzer rate in
Centaurus. Since this last requirement is impossible we deduce that
the Centaurus X-ray filaments are either not powered by conduction or
represent gas which is out of equilibrium, either cooling down (the
radiative cooling time is $\sim 10^8\yr$) or heating up, perhaps by
cool filament gas mixing in with the ambient hotter gas. 

\section*{Acknowledgments}

CSC and ACF thank the Royal Society for financial support, and we
thank Greg Taylor for producing the 330MHz image.  The archival HST
data presented in this paper were obtained from the Multimission
Archive at the Space Telescope Science Institute (MAST). STScI is
operated by the Association of Universities for Research in Astronomy,
Inc., under NASA contract NAS5-26555. Support for MAST for non-HST
data is provided by the NASA Office of Space Science via grant
NAG5-7584 and by other grants and contracts. The European Southern
Observatory (Chile) data were taken under programme ID 073.A-0077. The
Gemini Observatory data were obtained under programme ID
GS-2004A-Q-79. The Gemini telescopes are operated by the Association
of Universities for Research in Astronomy, Inc., under a cooperative
agreement with the NSF on behalf of the Gemini partnership: the
National Science Foundation (United States), the Particle Physics and
Astronomy Research Council (United Kingdom), the National Research
Council (Canada), CONICYT (Chile), the Australian Research Council
(Australia), CNPq (Brazil) and CONICET (Argentina).

{}

\end{document}